\documentclass[11pt, twocolumn, journal]{IEEEtran}
\usepackage{amsmath,epsfig,psfrag,subfig}
\usepackage{multirow}
\usepackage{amsfonts,color}
\usepackage{url}

\long\def\comment#1{}
\long\def\ignore#1{}
\long\def\lquote#1{} % \begin{quote} #1 \end{quote} }

\psfull
\title{Information Cascades in Feed-based Networks of Users with Limited Attention}

\author{\IEEEauthorblockN{Sameet Sreenivasan\IEEEauthorrefmark{1}\thanks{Research was in part sponsored by the Army Research Laboratory and was accomplished under Cooperative Agreement Number W911NF-09-2-0053 (Network Science CTA) and by the Office of Naval Research Grant No. N00014-09-1-0607. The views and conclusions contained in this document are those of the authors and should not be interpreted as representing the official policies, either expressed or implied, of the Army Research Laboratory or the U.S. Government. The U.S. Government is authorized to reproduce and distribute reprints for Government purposes notwithstanding any copyright notation here on.}, Kevin S. Chan\IEEEauthorrefmark{2}, Ananthram Swami\IEEEauthorrefmark{2}, Gyorgy Korniss\IEEEauthorrefmark{1} and Boleslaw Szymanski\IEEEauthorrefmark{1}\\}
\IEEEauthorblockA{\IEEEauthorrefmark{1}Rensselaer Polytechnic Institute, Troy, NY 12180\\ 
Email: {\{sameet, korniss, szymab\}}@rpi.edu}\\
\IEEEauthorblockA{\IEEEauthorrefmark{2}US Army Research Laboratory, 
Adelphi, MD 20783\\
Email: kevin.s.chan.civ@mail.mil, a.swami@ieee.org\\}}

\begin{document}
\maketitle
\begin{abstract}
%We build a model of information cascades on feed-based networks, taking into account the finite attention span of users, message generation rates and message forwarding rates. Using this model, we computationally measure the likelihood that an information cascade becomes viral (i.e. reaches a ``global'' scale, and how this likelihood depends on the extent of user attention, keeping other parameters of the model fixed. We simultaneously formulate an analytical approach to model the limited attention span, and show how the critical forwarding probabilities required to
%generate ``viral'' cascades, can be predicted given the extent of user-attention. We explore the parameter space of the model using extensive computational simulations on stylized networks. Further, we estimate the parameters of this model through data analysis of a Twitter data set. We find conditions for virality based on analysis of the branching factor in our dataset. 
We build a model of information cascades on feed-based networks, taking into account the finite attention span of users, message generation rates and message forwarding rates.
Using this model, we study through simulations, the effect of the extent of user attention on the probability that the cascade becomes viral.
In analogy with a branching process, we estimate the branching factor associated with the cascade process for different attention spans and different forwarding probabilities, and demonstrate that beyond a certain attention span, critical forwarding probabilities exist that constitute a threshold after which cascades can become viral. The critical forwarding probabilities have an inverse relationship with the attention span.
Next, we develop a semi-analytical approach for our model, that allows us determine the branching factor for given values of message generation rates, message forwarding rates and attention spans. The branching factors obtained using this analytical approach show good agreement with those obtained through simulations.
Finally, we analyze an event-specific dataset obtained from Twitter, and show that estimated branching factors correlate well with the cascade size distributions associated with distinct hashtags.
\end{abstract}
\begin{IEEEkeywords}
social computing, information cascades, limited attention
\end{IEEEkeywords}

\section{Modeling the effect of limited attention span on cascades in feed-based networks}
With the pervasiveness of social networking platforms, users are highly connected and have the ability to generate and forward information across networks. Feed-based networks such as Facebook's news feed, Twitter, and Instagram have become immensely popular, and using them users can scroll through streams of the latest information from sources that the user chooses to follow. It has become increasingly apparent that the volume of information being generated is far greater than the amount of information that users of these social networks can consume. Despite this preponderance of information, certain messages are able to ``go viral'', meaning that the message is forwarded and seen by most (or at least an asymptotically finite fraction) of the users in the social network. Conversely, there are many messages that are seen by few people and forwarded by no one. 

Given the preponderance of feed-based social networks, users will only look through a limited number of messages in their feed (before getting bored, fatigued, or interested in another topic or feed). If a particular message is deemed to be interesting, then the message is forwarded to the user's followers. As a consequence of the variety in the number of feeds one can monitor, individual messages end up competing for limited attention span resulting in a fat-tailed distribution of the popularity of individual messages \cite{Weng2012,gleeson2014competition}. 

Our approach is to define a model of networked user behavior within these feed-based networks to demonstrate the impact of the limited attention of users. Our model is based on three parameters that define the behavior of individual users: the probability of a user generating a new message, the probability that the user forwards a given message in his feed, and the length of the user feed. Simulations of our model support the idea that the dynamics of an information cascade follow a branching process \cite{Harris}, and how far the given branching process is above or below criticality depends on the depth of the feed to which a user typically devotes his attention. At low values of forwarding probability, the probability distribution of cascade sizes is a power-law with an exponential cutoff indicative of a sub-critical branching process while at large values of forwarding probability global cascades become increasingly prevalent. We analyze this model and the resulting branching process in detail, and also investigate a real dataset of tweets for signatures of the behavior suggested by our model.

A brief overview of related work is found in Section \ref{sec:background}. The model is introduced in Section \ref{sec:model} and the analysis is described in Section \ref{sec:analysis}. In Section \ref{sec:twitter}, we use a set of tweets to demonstrate that signatures of critical, sub-critical and supercritical behavior are also present in empirical data on cascades and therefore qualitatively support our model. 

\section{Background and Related Work} \label{sec:background}
In the past five years, the topic of information cascades has gathered the attention of both social network and network analysis researchers. In particular, recent theoretical work by Weng \cite{Weng2012} and Gleeson \cite{gleeson2014competition} have shown how simple models of message forwarding can give rise to a fat-tailed distribution for the degree of spread (popularity) of a given message. In Section \ref{sec:model}, we comment on how these models are similar to and different from the one we consider in this work.

Using attributes of the underlying graph as explanatory variables is a common approach to studying information cascades on networks. Typically, these networks have either directional links, through follower and followee relationships or bi-directional links, through friendship statuses.  Higham et al. \cite{higham14spikes} consider a model to predict spikes of activity while considering various graph-theoretic characteristics of these networks. In contrast, our analysis is based on abstracting the cascade as a branching process, and we apply this approach to data generated from simulations on scale-free networks, and to data collected from Twitter. 

Another approach common in studying information cascades in social networks is to consider community structure. Weng et al. \cite{weng13viral, weng14commstruct} investigate the impact of community structure on spreading of memes. The dynamics of these cascades are studied with respect to various complex contagion models. Others have used epidemic models to investigate the dynamics of information cascades. \cite{jin13rumors} considers these epidemic models with four states in epidemics and attempts to fit empirical data to parameterize these models. Analogs to states in this type of model are the user is \textit{susceptible} when open to viewing messages of a particular topic; \textit{exposed} while a message is in its queue; \textit{infected} after the message is forwarded, \textit{recovered} when it can return to the susceptible state. Our approach does not allow for such a recovered state.

We do not represent or study message content nor do we conduct sentiment analysis on the data. Modeling of these dynamics can provide more insight into the behavior of the users and global dynamics \cite{Doan2012sentiment}. Further, for tweets containing hashtags of interest, the tweet may represent negative or positive support of the topic, which may also provide additional dimensions of study. 

Finally, the model in this work does not consider placement of messages in feeds as conditions for virality. Hodas and Lerman \cite{hodas2012contagion, lerman2012contagion} study the impact of the placement of the messages in the queue and its impact on social contagion, where messages at the top of queues are more likely to be viewed. One can also consider other characteristics of the messages such as which other users forwarded the messages as well as their relationship to the current active user. In this work, our model is a FIFO stream.

A work related to biases in data collection, \cite{Morstatter13firehose} compares the full Twitter feed {\it Firehose} with the sampled {\it Gardenhose} Twitter Streaming API to which the majority of researchers have access.

\section{Model} \label{sec:model}
We abstract the social network under consideration as a graph $G$. For simplicity we assume the graph to be undirected, but results found using directed edges are qualitatively similar. We assume that the network degree distribution is scale-free as indicated by previous analyses of connectivity in social networks \cite{Kwak2010}
%Outward edges (from node $i$ to node $j$) indicates a follower relationship, where node $j$ follows node $i$. Inward edges indicate a followee relationship.
The parameters governing message propagation in our model are as follows:

\begin{itemize}
\item $p_n$: probability of new message generation
\item $p_r$: probability of message forwarding
\item $L$: length of message buffer
\end{itemize}

The dynamics of message cascading results from the following rules. In each time slot $t$, one user (node $i$) is randomly selected to become active, and then will choose to create a new message with probability $p_n$. If a new message is generated, then each of the followers of node $i$ will receive this message in their message feed, and, if needed, their oldest message is removed from their buffers. Then, node $i$ goes through its feed of up to $L$ messages one by one, starting from the most recent message. With probability $p_r$, each message will be independently forwarded to its neighbors. We assume that a node does not forward the same message twice. When a user receives a message that has been already forwarded, it will ignore this message although the message still occupies a place on its feed. 
The duration of a time slot is the time required for a node to go through the above set of $L+1$ actions. 

Given this model, we study how the popularity of a message -  the number of users forwarding that message - is statistically distributed. Note that in the rest of the paper, we interchangeable use the terms {\it queue} and {\it feed}.

Similar models have been studied in \cite{Weng2012} and \cite{gleeson2014competition}, but with the crucial difference that when a node is selected for an update, it either generates a new message or forwards a randomly chosen message from its queue. The former study has additional parameters in the model as well, making it less analytically tractable. In contrast the model in  \cite{gleeson2014competition} as shown therein, is analytically tractable; however the system approaches criticality only when the message generation probability tends to zero. Furthermore, a regime of supercritical behavior is not attainable. This is another point where our model contrasts with that of   \cite{gleeson2014competition} as we shall demonstrate in forthcoming sections.

\section{Results} \label{sec:twitter}
\subsection{Simulations} \label{sec:sims}
In this section, we describe the set of simulations conducted to test the model described in Section \ref{sec:model} and to gain understanding of the conditions under which one can expect global cascades, \textit{i.e.} a significant fraction of nodes forwarding the message.

We simulate the model on scale free networks consisting of $N=10000$ nodes with a power-law exponent of $\gamma = 2.5$ . Ten independent simulations were run, each for $T = 40 \times N = 4 \times 10^5$ timesteps. One timestep consists of one user being activated, potentially generating a new message, and going through its queue. To account for transient behavior of the queues populating, we begin collecting data on the number of users sharing specific messages only for those messages created after after $t = 10^5$ timesteps. Unless otherwise stated, we use $p_n = 0.45$. We refer to the number of users forwarding a given message as the {\it cascade size} associated with that message.

\begin{figure}[!ht]
    \subfloat[$L=10$\label{subfig:dist_L10}]{%
      \includegraphics[width=0.45\textwidth]{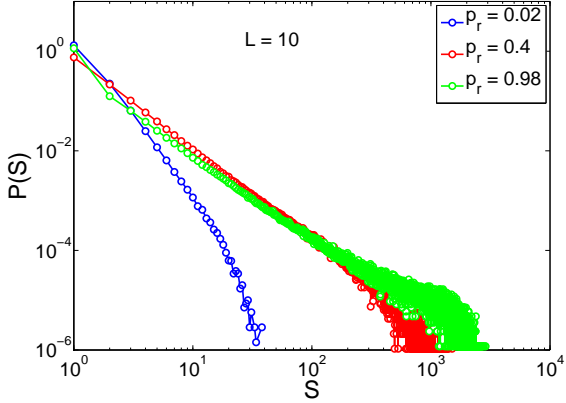}
    }
    \hfill
    \subfloat[$L=100$\label{subfig:dist_L100}]{%
      \includegraphics[width=0.45\textwidth]{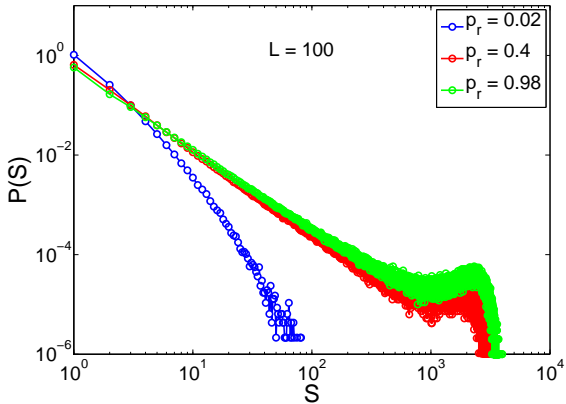}
    }
    \hfill
    \subfloat[$L=200$\label{subfig:dist_L200}]{%
      \includegraphics[width=0.45\textwidth]{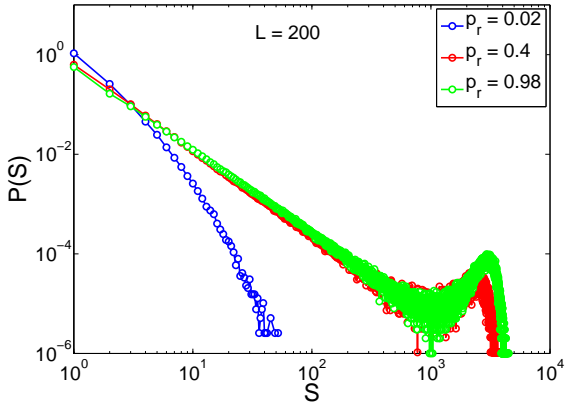}
    }
    \caption{Distribution of cascade size for various values of $L$}
    \label{fig:dist_L}
  \end{figure}

Figs \ref{subfig:dist_L10}-\ref{subfig:dist_L200} show simulation results for simulations of various values of forwarding probability $p_r$ and depth of the queue $L$. Observations are that for high forwarding probabilities, $p_r = 0.98$, global cascades are present for all $L$ shown. For low forwarding probabilities $p_r = 0.02$, the cascade sizes are distributed exponentially. For the intermediate forwarding probability shown $p_r = 0.4$, the cascade sizes transitions from power-law to exhibiting super-criticial behavior of global cascades.

In Fig. \ref{subfig:dist_L10}, for the lowest value of forwarding probability, cascade sizes are distributed exponentially, while at the intermediate value of $p_r = 0.4$, the distribution has a robust linear behavior (on the log-log plot) upto a specific cascade size, suggesting a power-law behavior of the cascade size distribution, modulated by an exponential cutoff. Finally, at the largest value of forwarding probability, the mass of the distribution appears to be increasingly shifting to the tail. The latter behavior becomes more pronounced as the queue length $L$ is increased - see Figs \ref{subfig:dist_L100} and \ref{subfig:dist_L200}. In general, the transitions in the form of the distribution are indicative of a branching process like behavior which many systems subject to cascades have been found to follow \cite{Goh2003,BeggsTimme2012}. Specifically, a branching process can be sub-critical, critical or super-critical, with the respective tree-sizes generated by the branching process being distributed exponentially, as a power-law, and bimodally with more and more of the distribution becoming concentrated at the higher mode.

Next, we estimate the critical forwarding probability $p_r$ at which the cascade distributions can be assumed to attain their power-law behavior as follows. The goal in general is to determine the ``branching factor" of the branching process i.e. the average number of copies that a typical node produces for any received message. When the branching factor is $1$, the branching process is said to be critical. The branching factor is estimated as follows::

%\begin{itemize}
%\item For a given message $m$, record all nodes that have received a copy of that message on their queue.
%\item For every recorded node for a specific message, $r_m$, count the number of copies of the message that the node produces. If the node never forwards the message, the number of copies is $0$. For a node that does share the message, the number of copies $n(r_m)$ is equal to the node degree minus one, since the neighbor from whom the node received the message will not forward it again.
%\item The branching factor is estimated as $\mu = \sum_m \sum_{r_m} n(r_m) /  \sum_m \sum_{r_m} 1$. 
%\end{itemize}

\begin{itemize}
\item For a given message $m$, lets put all nodes that have originated or received a copy of that message on their queue in the set $R_m$.
\item For every $r$ in set $R_m$ for a specific message $m$, count the number of copies of the message that the node produces. If the node never forwards the message, the number of copies is $n_m(r)=0$. For a node that does share the message, the number of copies $n_m(r)$ is equal to the node degree, if the node originates the message, or the node degree minus one if the node received the message, since the neighbor from whom the node received the message will not forward it again.
\item The branching factor is estimated as $\mu = \sum_m \sum_{r\in R_m} n_m(r) /  \sum_m \sum_{r\in R_m} 1$. 
\end{itemize}

\begin{figure}[!ht]
    \subfloat[Branching factor vs. Forwarding probability for various $L$. Horizontal line represents criticality point separating the subcritical and supercritical regimes.\label{subfig:branchfactorsim}]{%
      \includegraphics[width=0.45\textwidth]{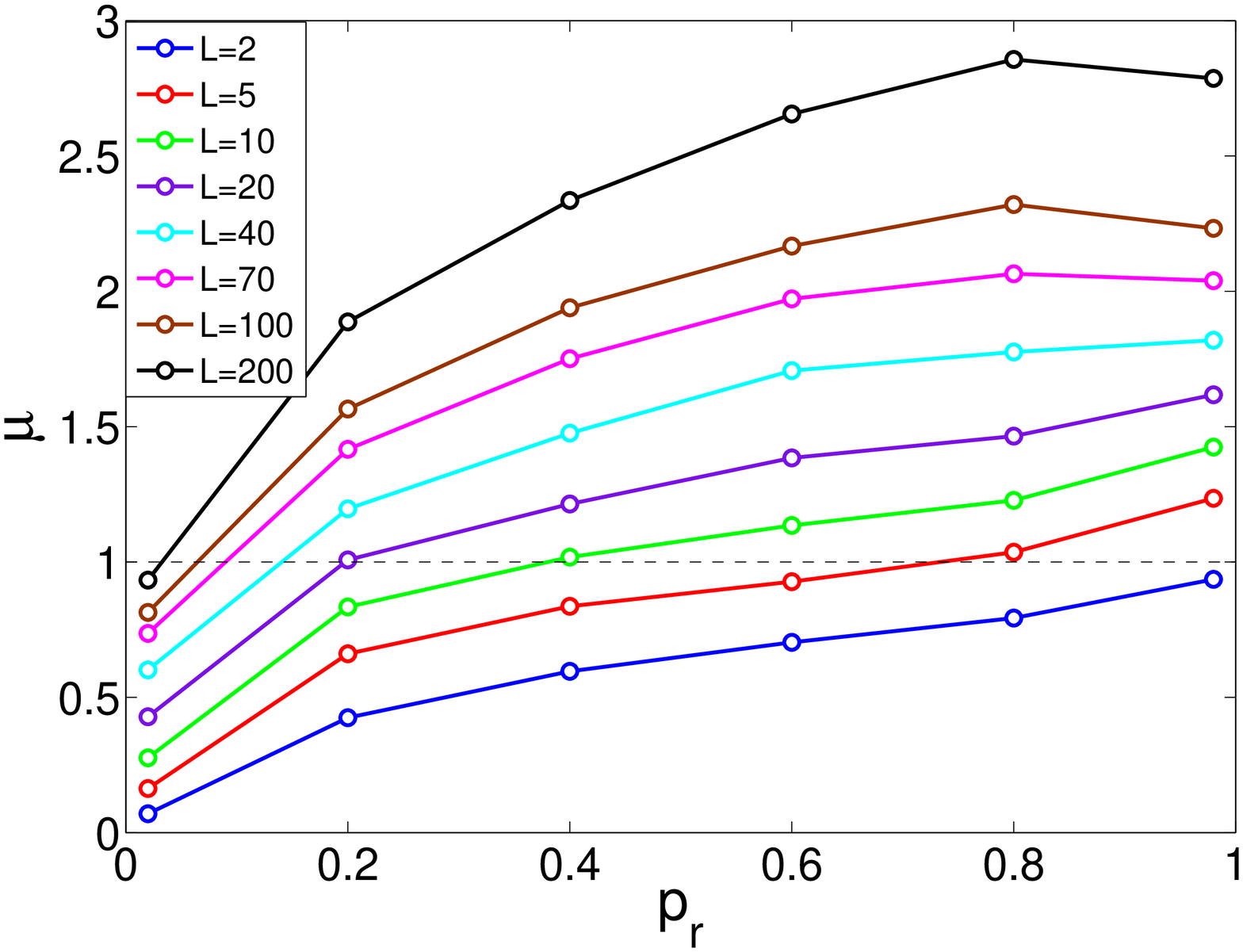}
    }
    \hfill
    \subfloat[Comparison of simulation and analysis of branching factor vs $p_r$ for various $L$.\label{subfig:branch_sim_vs_analysis}]{%
      \includegraphics[width=0.45\textwidth]{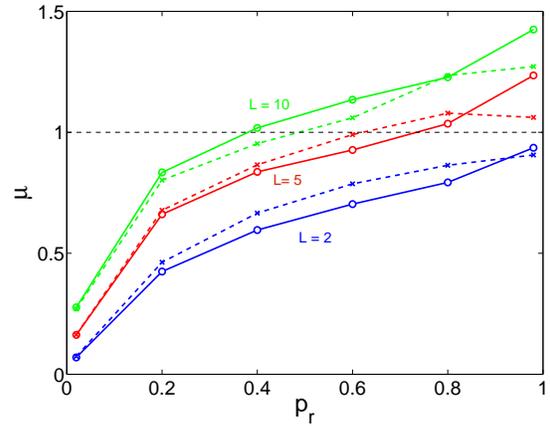}
    }
    \caption{Branching factor analysis for various values of $L$.}
    \label{fig:branching_factor}
  \end{figure}

The branching factor estimated in this manner is shown for a range of $p_r$ values for different queue lengths in Fig. \ref{subfig:branchfactorsim}. The dashed horizontal line indicates the critical branching factor. As a general trend, as $L$ increases, the $p_r$ at which the cascades approach a critical branching process decreases. Also of note is that for $L=2$, the dynamics of cascades will always operate in the subcritical regime. As $L$ tends to $\infty$, the system will increasingly operate in the supercritical regime. Fig. \ref{subfig:branch_sim_vs_analysis} shows the same analysis with simulation results superimposed to show the relative matching between the theory and simulation for three values of $L$.

In the next section, we derive an analytical estimate of the critical value of the forwarding probability $p_r$ at which the cascade distributions can be assumed to attain their power-law behavior thus making global cascades rare but possible.

\section{Analysis} \label{sec:analysis}
Here we describe our approach to analytically estimating the branching factor for the model given parameters $p_n$, $p_r$ and $L$. The basic idea is to estimate the mean number of copies of a message that a user produces given that it has received the message. For simplicity, and due to the low density of loops in randomly generated scale-free networks \cite{Newman2000}, we assume for purposes of analysis, that the structure on which the cascades are taking place is a tree with the same degree distribution as the original scale-free network. Further assuming that a node forwards messages only to its children on the tree ensures that any message is received by a node only once, and this vastly simplifies our analysis. 

In order to estimate the mean number of copies of a message produced by a node, we first estimate the effective forwarding probability of a received message by a node. A primary consideration here is the fact that the node can have multiple attempts at forwarding the message for as long as the message has not been pushed out of the queue by incoming messages. Thus the message gets some number of chances before exiting the user's window of attention. Let $\rho_n$ be the probability that the message survives on the node's queue for exactly $n$ rounds of the node's activity. Since in each round the message can be forwarded with a probability $p_r$, the effective forwarding probability for a given message can be written as:
\begin{equation}
p_f = \sum \rho_n (1-(1-p_r)^n)
\label{eff_fwd}
\end{equation}
since the term in parentheses on the right hand side represents the probability that the message is forwarded at least once in $n$ updates. 

The probability $\rho_n$ can in turn be derived from the survival probability $\mathcal{P}_n$ which captures the probability that the user receives less than $L$ messages in $n$ updates thus ensuring that the message under consideration remains on the user's queue for at least $n$ updates. The survival probability itself can be written as:
\begin{equation}
\mathcal{P}_n = \sum_j q_{j,n} u(m_j < L)
\label{survprob}
\end{equation}
where $q_{j,n}$ denotes the probability that $j$ events are forwarded to the user between $n$ updates of the user itself, and $u(m_j < L)$ is the probability that the number of messages received by the user as a result of these $j$ forwarding events is less than its queue length.

To obtain $q_{j,n}$, we treat the update of a node as a Poisson process with rate $1$. As a result, the process describing an event among any of the $k$ (incoming) neighbors of a node is a Poisson process with rate $k$. Utilizing these two independent Poisson process allows us to derive an expression for $q_{j,n}$. 

We obtain $u(m_j < L)$ by estimating the probability that a given number of messages are forwarded in $j$ neighborhood update events accounting for the fact that the forwarded messages could be newly generated ones (occurring with probability $p_n$ per update) or previously existing messages. In order to account for the fact that messages that have been forwarded once before are not forwarded again, we assume that each message in the queues of the neighborhood is eligible for forwarding with probability $\alpha$. To complete the calculation of  $u(m_j < L)$ we need to estimate $\alpha$ in a self-consistent manner. We do so by first deriving the probability that a message on a queue has age $w$ i.e. has survived for $w$ updates since it was received, and using it to compute the average probability that a message on a queue has not been forwarded yet. The former probability can be expressed in terms of $\alpha$, and the latter should be identical to $\alpha$. To solve this fixed-point equation, we scan for the value of $\alpha$ at which the self-consistency is satisfied. 

With $q_{j,n}$ and  $u(m_j < L)$ evaluated, we can estimate the survival probability $\mathcal{P}_n$, and from the latter derive $\rho_n = \mathcal{P}_{n}-\mathcal{P}_{n+1}$. Finally, using $\rho_n$, we can estimate the effective forwarding probability of a message. The number of copies of a message produced by a node is equal to the number of descendants that the node has in a tree. Thus, given the degree distribution $p_k$ of the network, we can derive the probability that the node produces $k$ copies of a message that was  forwarded to this node.

Following the arguments described above, we numerically evaluate the branching factor for given values of $p_n = 0.45$, $p_r$ and $L$. 

As a comparison of the simulation results and the analysis, we show the critical branching factor as a function of $p_r$ in Fig. \ref{subfig:branch_sim_vs_analysis}. There is close agreement for low to intermediate values of $p_r$. At higher values of $p_r$ and $L$, the true dynamics of cascades becomes incompatible (for small $L$ there is no divergence) with the consequences of the tree-like assumption in the analysis.

\section{Twitter Analysis} \label{sec:tweets}
As a complement to the analytical model and simulation results, we performed an empirical study of tweets from Twitter.

Approximately 3 million tweets were gathered, streaming a specific set of hashtags for the 2014 Superbowl (25 Jan 2014 to 4 Feb 2014, with the game held on 2 Feb 2014). The hashtags of interest were related to the Superbowl as well as commercial hashtags, all chosen before data collection. Per Twitter's convention, a forwarded message is called a retweet. Additionally, two periods of tweets were extracted, pregame and postgame tweets.

\subsection{Data Streaming Limitations}

\begin{figure}[!ht]
    \subfloat[Inter-event times for the Postgame dataset.\label{subfig:postgame_interevent}]{%
      \includegraphics[width=0.45\textwidth]{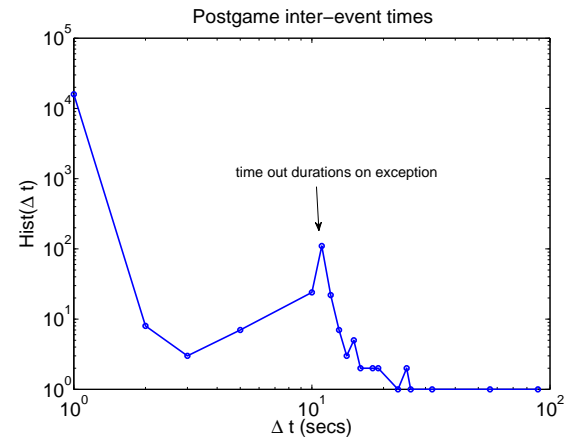}
    }
    \hfill
    \subfloat[In-degree distribution of mention network constructed from Postgame datset.\label{subfig:postgame_indegree}]{%
      \includegraphics[width=0.45\textwidth]{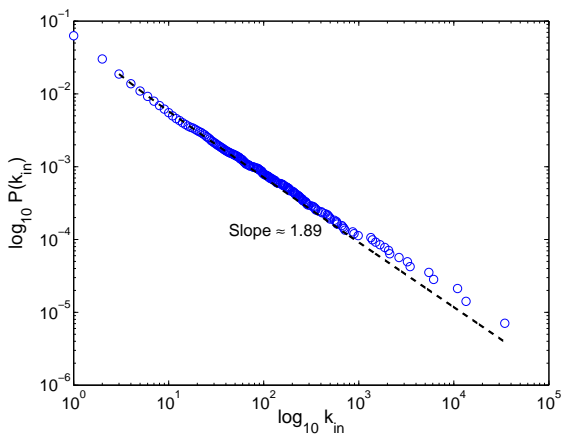}
    }
    \caption{Postgame dataset inter-event times and in-degree distribution.}
    \label{fig:postgame_dataset}
  \end{figure}

There are some inherent ``rate limits'' imposed on the collection of data from Twitter which make the data lossy \cite{Morstatter13firehose}. Also there are interruptions when there is a high rate of requests, and there are ``time-out'' periods to prohibit getting locked out. Fig. \ref{subfig:postgame_interevent} shows the time interval between successive tweets that we captured. Also ``time out'' periods are clearly visible. The spike in the middle of the data collection corresponds to the middle of the game in which Twitter traffic was significantly increased.

  \subsection{Followership network vs. Mention network}
Another limitation of the data is that we do not have network information (connectivity between users) as the gardenhose API does not have access to this information. With our collected data, we infer the implicit ``mention network'' between users, by assuming that if user $i$ mentioned user $j$ in a tweet, then $i$ is a follower of $j$  (i.e. $i$ points to $j$). The in-degree distribution of the inferred network is shown in Fig. \ref{subfig:postgame_indegree}. In our analysis, we focus on only the (undirected) giant component.

\subsection{Cascade size distributions}
\begin{figure}[!ht]
    \subfloat[Cascade size distributions of selected hashtags.\label{subfig:allcascades}]{%
      \includegraphics[width=0.45\textwidth]{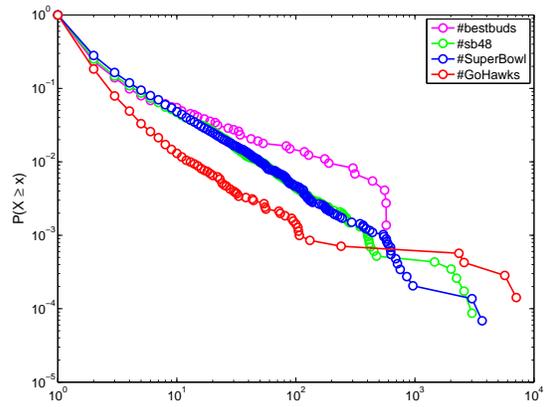}
    }
    \hfill
    \subfloat[Cascade size distributions with power-law maximum likelihood fits to \#GoHawks and \#Superbowl.\label{subfig:fits}]{%
      \includegraphics[width=0.45\textwidth]{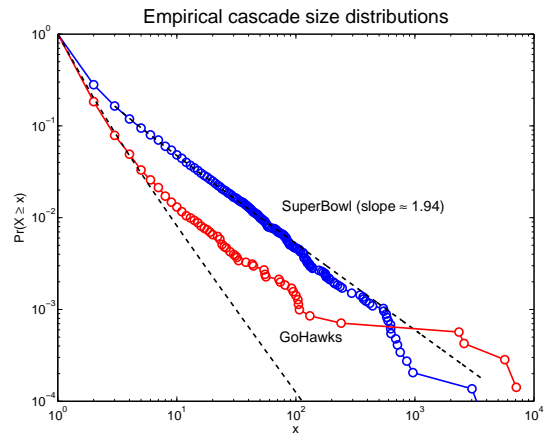}
    }
    \caption{Cascade size distributions for hashtags \#bestbuds, \#sb48, \#Superbowl and \#GoHawks.}
    \label{fig:cascades}
\end{figure}

Of the $31$ hashtags that we tracked (see Appendix), we selected four hashtags due to their relative abundance as compared to the rest. For each of these four hashtags, we gathered all tweets in our dataset that contained the hashtag under consideration, and none of the other hashtags that we were tracking. Within these tweets, we analyzed cascade sizes corresponding to each distinct tweet (each retweeted message has an identifier of the original message of which it is a copy). The cascade size distributions obtained through this analysis are shown in Fig.~\ref{subfig:allcascades}. A qualitative behavior similar to that seen in our simulations can also be observed here, although care must be taken in the comparison since we have shown the complementary cumulative distribution of cascade sizes here unlike in Figs \ref{subfig:dist_L10}-\ref{subfig:dist_L200}. However, signatures of subcritical, critical, and supercritical processes are observable here as well, with the hashtag {\it bestbuds} having a far smaller tail than that of the rest, while the hashtag {\it GoHawks} has a much fatter tail resembling a supercritical branching process. 

We further investigate the cascade size distributions associated with the two hashtags with the most data, {\it \#Superbowl} and {\it \# GoHawks}. A power-law distribution, as would be expected from a critical branching process, yields a good fit to the cascade size distribution of the former but results in a poor fit for the latter; see Fig. \ref{subfig:fits}. Note that it is the complementary cumulative distribution function for the cascade sizes to which we have fit the power-law. As we see later, the reason for the differences in the quality of the fits is the difference in the branching factors of the two hashtags. As mentioned earlier, for large branching factors, as is the case for {\it \#GoHawks}, the process is supercritical, and therefore results in the power-law fit being poor.

\subsection{Branching Factor Analysis}
Next, we estimate the branching factor for the cascades. Due to the limitations of data, we use the following approach:

For a given message $x$:
\begin{enumerate}
\item Record all nodes that have retweeted a copy of that message. Count the number of copies they generate i.e. the number of their followers (on the mention network).
\item We assume that the set of recorded nodes and their combined neighborhood comprise the set of nodes receiving $x$. Call this set the ``touched'' set.
\item Branching factor = total number of copies/ Size of touched set. 
\end{enumerate}

The order of hashtags according to branching factor in Table \ref{tab:branching} is consistent with ordering based on the fatness of the tails of their cascade size distributions in Figure \ref{subfig:fits}.

\begin{table}[htp]
\begin{center}
\begin{tabular}{|c|c|}
\hline
Hashtag & Estimated Branching Factor \\
\hline
bestbuds & 1.097 \\
Superbowl & 1.756 \\
sb48 & 5.310 \\
GoHawks & 12.135 \\
\hline
\end{tabular}
\end{center}
\caption{Branching factors for several hashtags}
\label{tab:branching}
\end{table}

\begin{figure}[!ht]
    \subfloat[\#SuperBowl \label{subfig:superbowl}]{%
      \includegraphics[width=0.45\textwidth]{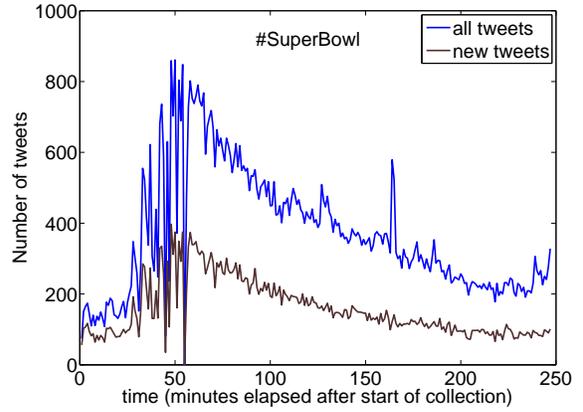}
    }
    \hfill
    \subfloat[\#GoHawks \label{subfig:gohawks}]{%
      \includegraphics[width=0.45\textwidth]{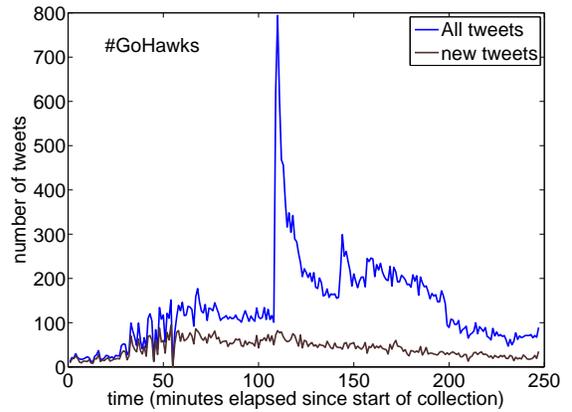}
    }
    \caption{Activity of new tweets and retweets for \#SuperBowl and \#GoHawks over time.}
    \label{fig:activity}
\end{figure}

We studied tweets for \#Superbowl and \#GoHawks, looking at the retweet distributions over time. This is shown in Figures \ref{subfig:superbowl} and \ref{subfig:gohawks}, where the retweets are shown as a subset of the new tweets over time. Additionally, Tables \ref{tab:superbowl10} and \ref{tab:gohawks10} show an excellent match between actual sources of tweet and the first tweeter of tweet. These tables show the 10 most widely spread tweets with \#SuperBowl and \#GoHawks. Of note are the boldfaced rows, which indicate that the original tweet was captured in the streamed dataset.

\begin{table}[htp]
\begin{center}
\begin{tabular}{|c|c|c|}
\hline
Rank & Follower count (original) & Follower count (dataset) \\
\hline
1 & 1113326 & 263 \\
2 & 30303 & 457 \\
3 & 4088 & 631 \\
4 & 20096 & 6 \\
5 & 557775 & 171 \\
6 & 874 & 681 \\
7 & 481634 & 46 \\
8 & \textbf{251500} & \textbf{251500} \\
9 & 24777739 & 200 \\
10 & 159544 & 538 \\
\hline
\end{tabular}
\end{center}
\caption{Top retweeted \#SuperBowl tweets and comparison of original tweeters and tweeters in dataset. \textbf{Bold} entries indicate that the original source was captured in the streamed dataset.}
\label{tab:superbowl10}
\end{table}

\begin{table}[htp]
\begin{center}
\begin{tabular}{|c|c|c|}
\hline
Rank & Follower count (original) & Follower count (dataset) \\
\hline
1 & \textbf{650504} & \textbf{650504} \\
2 & 467808 & 20730\\
3 & 54370 & 5 \\
4 & \textbf{898024} & \textbf{898204} \\
5 & \textbf{108} & \textbf{108} \\
6 & 453854 & 54 \\
7 & 642489 & 232 \\
8 & \textbf{244} & \textbf{244}\\
9 & \textbf{6220} & \textbf{6220} \\
10 & 440624 & 165 \\
\hline
\end{tabular}
\end{center}
\caption{Top retweeted \#GoHawks tweets and comparison of original tweeters and tweeters in dataset. \textbf{Bold} entries indicate that the original source was captured in the streamed dataset.}
\label{tab:gohawks10}
\end{table}

\section{Discussion}
We have presented and analyzed here a parsimonious model for cascades in feed-based social network environments. Using this model. we show that the cascading behavior can fall into three regimes, corresponding to the analogous regimes of a branching process. The particular regime that the cascading behavior falls into depends on the combination of parameters in the model. What differentiates our results from prior theoretical results \cite{gleeson2014competition} is that supercritical behavior is possible in our model. Furthermore, empirical data obtained from Twitter also qualitatively supports our hypothesis that cascading behavior can fall into three regimes. 

%\clearpage
\section{Appendix}

List of hashtags gathered from Twitter

\begin{table}[htp]
\begin{center}
\begin{tabular}{|c|c|}
\hline
GoodToBeBad & UpForWhatever \\
CupTherapy & NoRoomForBoring \\
FuelYourPleasure & KissForPeace \\
itsgotime& HowMatters \\
Halftime& AmericaIsBeautiful \\
sorrycokeandpepsi& StayUncompromised \\
VW& Wings \\
sb48 & Covered \\
Uncovered & bestbuds \\
salute & SuperBowl \\
12thman & GoHawks \\
SeaHawks &  beastmode \\
tgibf & youmadbro \\
teamsmallbiz &  unitedinorange \\
broncospride & peyton \\
nfl&  \\
\hline
\end{tabular}
\end{center}
\caption{List of hashtags collected from Twitter stream}
\label{tab:hashtags}
\end{table}

\bibliographystyle{IEEEtran}
\bibliography{twitter}
\end{document}